\documentclass[preprint,12pt]{elsarticle}

\usepackage{amssymb}
\usepackage{algorithm}
\usepackage{booktabs}
\usepackage{csquotes}
\usepackage{hyperref}
\usepackage{lineno}
\usepackage{listings}
\usepackage{multicol}
\usepackage{multirow}
\usepackage{optidef}
\usepackage[sfdefault]{FiraSans}
\usepackage{tabularx}
\usepackage{url}
\usepackage{xspace}
\usepackage[usenames,dvipsnames]{color}
\usepackage{colortbl}
\usepackage{stackengine}

\definecolor{light-gray}{gray}{0.95}
\definecolor{pgreen}{RGB}{5,205,107}
\definecolor{pblue}{RGB}{2,154,223}
\definecolor{pteal}{RGB}{0,128,128}
\definecolor{firebrick}{rgb}{0.7, 0.13, 0.13}

\usepackage[framemethod=tikz]{mdframed}
\mdfdefinestyle{mpdframe}{
    frametitlerule              =true,
    roundcorner                 =1pt,
    middlelinewidth             =1pt,
    innermargin                 =0.1cm,
    outermargin                 =0.1cm,
    innerleftmargin             =0.1cm,
    innerrightmargin            =0.1cm,
    innertopmargin              =0.1cm,
    innerbottommargin           =0.1cm,
}

\def\BibTeX{{\rm B\kern-.05em{\sc i\kern-.025em b}\kern-.08em
T\kern-.1667em\lower.7ex\hbox{E}\kern-.125emX}}

\lstdefinelanguage[programming]{TeX}[AlLaTeX]{TeX}{%
  deletetexcs={title,author,bibliography},%
  deletekeywords={tabular},
  morekeywords={abstract},%
  moretexcs={chapter},%
  moretexcs=[2]{title,author,subtitle,keywords,maketitle,titlerunning,authorinfo,affiliation,authorrunning,paperdetails,acks,email},
  moretexcs=[3]{addbibresource,printbibliography,bibliography},%
}%
\newcommand{\proDJ}[0]{{\textsc{ProDJ}}\xspace}
\newcommand*{\CTAN}[1]{\href{http://ctan.org/tex-archive/#1}{\nolinkurl{CTAN:#1}}}

\newcommand{\graphhopper}[0]{GraphHopper\xspace}
\newcommand{\pdfbox}[0]{PDFBox\xspace}
\newcommand{\broadleaf}[0]{BroadLeaf\xspace}
\newcommand{\gephi}[0]{Gephi\xspace}
\newcommand{\revisedjssold}{\textcolor{black}}
\newcommand{\revisedjss}{\textcolor{black}}
\newcommand{\revisedjssfinal}{\textcolor{black}}

\newcommand{\formattedquote}[1]{%
\begingroup
\setlength{\tabcolsep}{12pt}
\renewcommand*{\arraystretch}{1.8}
\addstackgap[4pt]{%
\begin{tabular}{!{\color{violet}\vrule width 2 pt}p{12cm}}
\emph{#1}\\
\end{tabular}%
}
\endgroup
}

\journal{Journal of Systems and Software}

\begin{document}

\begin{frontmatter}

\title{Serializing Java Objects in Plain Code}

\author[inst1]{Julian Wachter}
\affiliation[inst1]{organization={Karlsruhe Institute of Technology},country={Germany}}

\author[inst2]{Deepika Tiwari}
\author[inst2]{Martin Monperrus}
\author[inst2]{Benoit Baudry}

\affiliation[inst2]{organization={KTH Royal Institute of Technology},country={Sweden}}

\begin{abstract}
In managed languages, serialization of objects is typically done in bespoke binary formats such as Protobuf, or markup languages such as XML or JSON.
The major limitation of these formats is readability.
Human developers cannot read binary code, and in most cases, suffer from the syntax of XML or JSON.
This is a major issue when objects are meant to be embedded and read in source code, such as in test cases.

To address this problem, we propose plain-code serialization.
Our core idea is to serialize objects observed at runtime in the native syntax of a programming language.
We realize this vision in the context of Java, and demonstrate a prototype which serializes Java objects to Java source code.
The resulting source faithfully reconstructs the objects seen at runtime.
Our prototype is called \proDJ and is publicly available. 
We experiment with \proDJ to successfully plain-code serialize $174,699$ objects observed during the execution of 4 open-source Java applications.
Our performance measurement shows that the performance impact is not noticeable.
\revisedjssold{Through a user study, we demonstrate that developers prefer plain-code serialized objects within automatically generated tests over their representations as XML or JSON.}
\end{abstract}


\end{frontmatter}


\section{Introduction}\label{sec:introduction}
Serialization consists in representing runtime data in a format that can be transmitted \cite{godefroid2020intelligent} or stored \cite{liu2020native}.
Developers serialize objects either through native programming language support, or with third-party libraries that support serialization of objects in different formats \cite{viotti2022survey}. 
The serialization format may be binary, such as Protocol Buffers \cite{protobuffs, eddelbuettel2014rprotobuf}, or use a markup language such as JSON \cite{harrand2021behavioral} or XML \cite{bray1997extensible}.

Serialization has many use cases, such as sharing data among services that are external \cite{neumann2018analysis} or internal \cite{pourhabibi2020optimus} to a software system.
Serialized objects can also be used within tests, to reproduce failures \cite{artzi2008recrash}, or to initialize test inputs to a target state \cite{KimXSO22, pankti, alshahwan2024observation}.

The problem space of object serialization has three dimensions: correctness, performance, and readability.
A few studies have focused on the correctness of serialization techniques \cite{viotti2022survey}, including assessing the behavior of serialization tools through formal proofs  \cite{ye2019verified}, or back-to-back testing \cite{harrand2021behavioral}.
Performance is a major challenge for serialization, and several works have focused on optimizing it in the context of wireless communication \cite{wong2004xstream}, microservices data exchange \cite{pourhabibi2020optimus}, or efficient test execution \cite{gligoric2011codese}.

However, readability is an aspect of serialization that has been neglected in previous work.
Producing readable serialized objects is the problem we address in this paper.
This is of utmost importance when serialized objects are consumed by humans. 
\revisedjss{For example, when serialized objects are included in test cases as part of test inputs \cite{artzi2008recrash,KimXSO22,pankti, alshahwan2024observation}, readability is essential to let developers  understand the tests.}
Clearly, mainstream serialization techniques fall short with respect to readability.
For example, \autoref{lst:introduction-test-with-xml} shows an example of a test case generated by the automated test generation tool Pankti \cite{pankti}, a tool that creates tests from objects at runtime.
In \autoref{lst:introduction-test-with-xml}, the test inputs are in the XML format, as is the case for the data used to populate the \texttt{receiver} and \texttt{key} variables.
XML strings arguably makes the code hard to read, making it challenging to understand the test's intention.

\begin{lstlisting}[
  language=Java,
  caption={Serialized objects in XML are not meant for human consumption, they are not readable. In this paper, we aim to serialize objects in plain code instead, as demonstrated later in \autoref{lst:introduction-test-plain-code} and \autoref{lst:prodj-test}},
  label={lst:introduction-test-with-xml},
  float,
  numbers=left,
  morekeywords={enum}
]
@Test
public void testGetNameAsString() {
  // Arrange
  COSDictionary receiver = deserializeFromXML("""
  <org.apache.pdfbox.cos.COSDictionary>
    <direct>false</direct>
    <needToBeUpdated>false</needToBeUpdated>
    <items class=\"org.apache.pdfbox.util.SmallMap\">
      <mapArr>
        <org.apache.pdfbox.cos.COSName>
          <direct>false</direct>
          <name>BaseFont</name>
          <hashCode>-1657267488</hashCode>
        </org.apache.pdfbox.cos.COSName>
        <org.apache.pdfbox.cos.COSName>
          <direct>false</direct>
          <name>Times-Roman</name>
          <hashCode>1058166902</hashCode>
        </org.apache.pdfbox.cos.COSName>
      </mapArr>
    </items>
  </org.apache.pdfbox.cos.COSDictionary>
  """);
  COSName key = deserializeFromXML("""
  <org.apache.pdfbox.cos.COSName>
    <direct>false</direct>
    <name>BaseFont</name>
    <hashCode>-1657267488</hashCode>
  </org.apache.pdfbox.cos.COSName>
  """);
  // Act
  String actual = receiver.getNameAsString(key);
  // Assert
  assertThat(actual).isEqualTo("Times-Roman");
}
\end{lstlisting}

\begin{lstlisting}[
  language=Java,
  caption={\revisedjssold{Automatically generated test containing plain-code serialized runtime objects \texttt{receiver} and \texttt{key}. Compared to the XML object representations in \autoref{lst:introduction-test-with-xml}, this test is arguably more readable and maintainable.}},
  label={lst:introduction-test-plain-code},
  float,
  numbers=left,
  morekeywords={enum}
]
@Test
@DisplayName("getNameAsString(org.apache.pdfbox.cos.COSName)")
public void testGetNameAsString_ProDJ() {
  // Arrange
  COSDictionary receiver = this.createCOSDictionary();
  COSName key = COSName.BASE_FONT;
  // Act
  String actual = receiver.getNameAsString(key);
  // Assert
  assertThat(actual).isEqualTo("Times-Roman");
}

COSDictionary createCOSDictionary1() {
  // Recreated from trace
  COSDictionary receiver = new COSDictionary();
  COSName key = COSName.TYPE;
  COSBase value = COSName.FONT;
  receiver.setItem(key, value);
  COSName key0 = COSName.SUBTYPE;
  COSBase value0 = COSName.TYPE1;
  receiver.setItem(key0, value0);
  COSName key1 = COSName.BASE_FONT;
  receiver.setName(key1, "Times-Roman");
  COSName key2 = COSName.ENCODING;
  COSBase value1 = COSName.WIN_ANSI_ENCODING;
  receiver.setItem(key2, value1);
  return receiver;
}
\end{lstlisting}

\revisedjss{To solve the problem of unreadable serialized data, such as JSON/XML test fixtures, we propose the novel concept of \emph{plain-code serialization}, which consists in serializing an object directly into code statements.}
By serializing objects to plain code, the serialized representation becomes readable, using a syntax that is both concise and natural to developers.
\revisedjssold{We illustrate this with the semantically equivalent test of \autoref{lst:introduction-test-plain-code}, which contains the same serialized objects as \autoref{lst:introduction-test-with-xml} represented as plain code, instead of XML.}
In addition, the compiler gives us strong guarantees about the static correctness of the serialized objects. 
Deserialization then simply means executing the code.
In the context of tests, this completely removes  dirty data like XML strings, thus making the generated tests more readable and self-contained \cite{afshan2013evolving}.  

In this paper, we propose two algorithms for plain-code serialization.
First, our structure-based serialization strategy relies on static analysis of the source code to identify a minimal sequence of constructor and setter calls, that are able to initialize an object to a desired state.
Second, the trace-based serialization strategy records constructor and method invocations and subsequently generates a corresponding plain-code version that replays these invocations to construct and initialize an object. 
We prototype the concept in Java, in a tool called \proDJ that serializes JVM objects in plain code.

We demonstrate the feasibility of plain-code serialization by evaluating \proDJ with $4$ well-known Java projects.
\proDJ successfully serializes $174,699$ Java objects observed at runtime, producing Java code that can reconstruct them.
We find that the objects are faithfully reconstructed through a full-cycle of plain-code serialization and deserialization.
\revisedjssold{Through a user study with 17 developers, we also demonstrate that the use of plain-code serialized objects contributes to the readability of automatically generated tests, relative to XML- or JSON-serialized objects.}

We summarize our contributions as follows:
\begin{itemize}
    \item The concept of plain-code serialization: producing source code statements that reconstruct an object to a target state, contrasting with binary (e.g. Protobuf) or textual format (e.g. JSON) serialization. 

    \item A publicly available prototype implementation of \proDJ for future research on advanced serialization in the JVM\footnote{\url{https://github.com/ASSERT-KTH/prodj}}.
    
    \item An evaluation of \proDJ on $4$ notable open-source projects in Java, involving $174,699$ serialized objects, assessing the feasibility of  plain-code serialization.

\end{itemize}
\proDJ and our experimental data are available in our replication package\footnote{\url{https://doi.org/10.5281/zenodo.7902538}}. 

\section{Plain-code Serialization}\label{sec:contribution}

In this section, we describe our key conceptual contribution: serializing runtime objects in plain code.
This is done by monitoring an application as it executes, capturing objects, and translating them to plain Java code.

\subsection{Concept \& Overview}\label{sec:concept}

Plain-code serialization means representing a runtime object with the concrete syntax of a programming language.
In this context,
\emph{serialization} means producing code from an object, such as a sequence of statements, that can be compiled and executed.
\emph{Deserialization} means executing the  plain-code representation of the runtime object, an operation we call in this paper \emph{object reconstruction}. 

The reconstruction operation for any object $o$ must conform to the invariant presented in \autoref{eq:object-reconstruction}.
\begin{equation} \label{eq:object-reconstruction}
\forall o, reconstruct(o) \equiv o
\end{equation}
\begin{equation} 
\label{eq:reconstructdef}
reconstruct(x) = evaluate(plain\_code\_serialize(x))
\end{equation}

The equivalence relationship ($\equiv$) means that the object after reconstruction is in the same state as the one observed at runtime. 
While \autoref{eq:object-reconstruction} applies to any serialization library, \autoref{eq:reconstructdef} is specific to plain-code serialization, with $reconstruct$ being the evaluation of the plain-code serialized representation of $x$ obtained from the function $plain\_\allowbreak{}code\_\allowbreak{}serialize$.

\emph{Benefits.}
\revisedjssold{Plain code serialization has two major benefits: readability and correctness checking.
First, the serialized objects are readable for developers,  using the programming language syntax for which they have been trained for.
Second, when serializing to a statically typed language, plain-code serialization provides guarantees about the serialized data.
In this case, the compiler checks for the validity of the object with the full power of the type system, capturing domain-specific correctness that one typically does not have with binary or markup serialization.}

In this paper, we devise two complementary strategies to map runtime objects to plain code:
structure-based serialization and trace-based serialization.
For structure-based object serialization, we statically analyze a class to identify all possible ways for setting each field of the object, such as using constructors or setters.
However, an object may not be fully reconstructible using the structure-based strategy.
This is due to the presence of internal, mutable state in object-oriented programs that are not exposed through setters. 
The second strategy, trace-based object serialization, overcomes this limitation.
It is based on recording a sequence of events that modify the state of an object  at runtime.
This recorded sequence of constructors and mutating methods, is then used to reconstruct the object to its target state.

We now illustrate how these two strategies work and complement each other. 
Consider the \texttt{Monkey} and \texttt{Habitat} classes  in \autoref{lst:extraction}.
\texttt{Monkey} is serializable using the structure-based strategy.
The three possible reconstruction plans for the \texttt{Monkey} class are presented in 
\autoref{lst:reconstruction-plans}.
To create the Java code for a given \texttt{Monkey} object, the \emph{\$age\$}, \emph{\$eyeColor\$}, and \emph{\$habitat\$} meta-variables are replaced with actual values from runtime.
However, note that the \texttt{Habitat} class in \autoref{lst:extraction} contains the private field \texttt{area}.
This field is part of the internal state of \texttt{Habitat} and, according to the information hiding principle, not directly exposed.
Therefore, we use trace-based object serialization for \texttt{Habitat}.
For example, recording the constructor invocation of \texttt{Habitat}, as well as an invocation of its \texttt{grow} method would allow us to reconstruct a \texttt{Habitat} object with coordinates \texttt{"{}42, 42{}"} and an \texttt{area} of \texttt{2.0}.
In the following subsections, we present the structure-based and trace-based object reconstruction strategies in detail.

\begin{lstlisting}[
  language=Java,
  caption={The example \texttt{Monkey} class to be serialized, and its associated types},
  label={lst:extraction},
  float,
  numbers=left,
  morekeywords={enum}
]
public class Monkey {
  private int age; 
  public Color eyeColor; 
  public Habitat habitat;

  public Monkey(int age) { this.age = age; }
  
  public Monkey(int age, Color eyeColor, Habitat habitat) {
    this(age);
    this.eyeColor = eyeColor;
    this.habitat = habitat;
  }

  public void setEyeColor(Color eyeColor) { this.eyeColor = eyeColor; }

  public String eatBanana(int dozens) {
    int quantity = dozens * 12;
    return "Eating " + quantity + " bananas";
  }
}

public enum Color { RED, GREEN, BLUE; }

public class Habitat {
  private String coordinates;
  private double area;

  public Habitat(String coordinates) { this.coordinates = coordinates; }

  public void grow(double newArea) {
    double newValue = this.area + newArea;
    this.area = newValue;
  }
}
\end{lstlisting}

\subsection{Structure-based Serialization (\textsc{sb})}\label{sec:structure-based-strategy}

The core idea of the structure-based object serialization strategy is to statically analyze a class and build a reconstruction plan.
\revisedjss{This is done by searching the class for statement which set the value of fields, such as constructor or setter method calls.}
The plan is instantiated at runtime for a given object.

\subsubsection{Synthesizing Reconstruction Plans}
A \emph{\textbf{reconstruction plan}} for a class is a template with actions that express how an object of the class is instantiated.
The plan contains meta-variables which are populated with the values observed at runtime.
When instantiated with a captured runtime object, the reconstruction plan is a complete, executable code snippet.
\autoref{eq:SB} presents the formulation of object reconstruction with structure-based serialization, as a specialization of idiomatic object reconstruction presented in \autoref{eq:object-reconstruction}. 
\begin{equation} \label{eq:SB}
evaluate(instantiated\_reconstruction\_plan(o)) \equiv o
\end{equation}

An \textbf{\emph{action}} within a reconstruction plan is a statement that sets zero or more fields within an object.
The $5$ action types used for structure-based object reconstruction are denoted as \textsc{sb} in \autoref{tab:action-types}.
Each action can be constructing ($\alpha$) or field-setting ($\beta_f$ sets a field $f$).
Constructor calls and calls to factory methods are constructing actions, whereas field assignments and setter calls are field-setting actions.
\revisedjssold{We analyze a target class to extract all possible actions that can be combined in different ways to automatically synthesize reconstruction plans.}
Each action has one or more meta-variables.
A \emph{\textbf{meta-variable}} is a placeholder for a field that is being set through the action.
When a plan is instantiated to serialize a runtime object, each meta-variable is replaced with the actual value of the field.

\begin{table*}
\centering
\begin{tabularx}{\textwidth}{rcXl}
\toprule
\textbf{\textsc{Action Type}} & \textbf{\textsc{Category}} & \textbf{\textsc{Explanation}} & \textbf{\textsc{Used by}}\\
\midrule
\noalign{\global\arrayrulewidth=0.5mm}
call constructor & $\alpha$ & constructs an instance using a constructor, potentially with parameters & \textsc{sb}, \textsc{tb}\\
\arrayrulecolor{light-gray}\hline
call factory method & $\alpha$ & creates an instance by calling a factory method & \textsc{sb}\\
\arrayrulecolor{light-gray}\hline
use enum constant & $\alpha$ & uses an enum instance identified by its name & \textsc{sb}\\
\arrayrulecolor{light-gray}\hline
call method & $\beta_f$ & sets field $f$ by calling a method with parameters & \textsc{sb}, \textsc{tb}\\
\arrayrulecolor{light-gray}\hline
assign field & $\beta_f$ & sets field $f$ by directly assigning it a value & \textsc{sb}\\
\arrayrulecolor{light-gray}\hline
use standard charset & $\alpha$ & creates a Java charset by using one of the constants in \texttt{StandardCharsets} & \textsc{tb}\\
\arrayrulecolor{light-gray}\hline
use static field & $\alpha$ & uses a public static field whose value is reference equal to the object under construction & \textsc{tb}\\
\arrayrulecolor{light-gray}\hline
use object reference & $\alpha$ & saves the object id, the object is then reconstructed using the strategy in \autoref{sec:trace-based-strategy} & \textsc{tb}\\
\end{tabularx}
\caption{The different types of actions for plain-code serialization, and the categories they belong to. $\alpha$ denotes the set of constructing actions, $\beta_f$ the set of actions that set field $f$. \textsc{sb} refers to structure-based object reconstruction, \textsc{tb} to trace-based object reconstruction.}
\label{tab:action-types}
\end{table*}

Let us illustrate the generation of a reconstruction plan using the \texttt{Monkey} class in \autoref{lst:extraction}.
A \texttt{Monkey} has an integer value for \texttt{age}, and an \texttt{eyeColor}, which takes possible values from an \texttt{enum}.
The \texttt{Monkey} also has a \texttt{Habitat}.
Each field of \texttt{Monkey} has a corresponding meta-variable, namely \emph{\$age\$}, \emph{\$eyeColor\$}, and \emph{\$habitat\$}.
There are three possible reconstruction plans for \texttt{Monkey}, which we present in \autoref{lst:reconstruction-plans}.
The most idiomatic way to create a \texttt{Monkey} is using the first plan, which  contains a single constructing action.
It creates a \texttt{Monkey} by calling the three-argument \texttt{age}, \texttt{color}, and \texttt{habitat} constructor.
The second and third plans consist of one constructing and two field-setting actions.
The second plan calls the \texttt{age} constructor, followed by the \texttt{eyeColor} setter, and finally directly assigns the \texttt{habitat} field.
The last plan has actions to call the \texttt{age} constructor, and directly assign both the \texttt{eyeColor} and \texttt{habitat} fields.

\begin{lstlisting}[
  language=Java,
  caption=Three possible reconstruction plans for \texttt{Monkey},
  label={lst:reconstruction-plans},
  float,
  numbers=left,
  morekeywords={enum}
]
// Reconstruction plan 1
Monkey monkey = new Monkey(%*\emph{\$age\$}*), %*\emph{\$eyeColor\$}*), %*\emph{\$habitat\$}*));

// Reconstruction plan 2
Monkey monkey = new Monkey(%*\emph{\$age\$}*));
monkey.setEyeColor(%*\emph{\$eyeColor\$}*));
monkey.habitat = %*\emph{\$habitat\$}*);

// Reconstruction plan 3
Monkey monkey = new Monkey(%*\emph{\$age\$}*));
monkey.eyeColor = %*\emph{\$eyeColor\$}*);
monkey.habitat = %*\emph{\$habitat\$}*);
\end{lstlisting}

\subsubsection{Optimizing Readability}
As is the case for \texttt{Monkey}, we enumerate all possible reconstruction plans for a target class.
However, we are interested in the shortest and most idiomatic plan, to reconstruct an object using the minimum number of statements \cite{bach2020determining}.
To quantify how idiomatic a reconstruction plan is, we first assign each action a pre-defined integer cost value.
For example, calling a constructor has a very low cost, while directly setting a field has a higher cost.
We then select the reconstruction plan with the lowest cumulative cost.
We formulate this optimization as  a constraint problem, shown in 
\autoref{prob:structure-based}.
\revisedjss{In the constructed problem, every action $i$ is assigned a boolean selection variable $a_i$.
If the model value for $a_i$ is true, the action is selected.
If it is false, the action will not be part of the reconstruction plan.}
If an action sets a field $f$, it is added to the set $\beta_f$, and if it is a constructing action, also to $\alpha$.

The cost of an action $i$ is denoted by $cost_{i}$ and the global optimization objective is modeled with \autoref{prob:structure-based-goal}.
\autoref{prob:structure-based1} ensures that at least one constructing action is selected; \autoref{prob:structure-based2} ensures that there is a single constructing action; \revisedjssold{and \autoref{prob:structure-based3} means that all fields are assigned at least once.}
If we find a solution to the constraint problem, all used actions are deterministically combined to produce a reconstruction plan.
\revisedjssold{The plan will be also applied repeatedly for recursive structures.}
%
\begin{mini!}|s|[1]
{a \in \{0,1\}^n} 
{\sum_{i=1}^n a_i \cdot cost_{i}\protect\label{prob:structure-based-goal}} 
{\label{prob:structure-based}} 
{} 
\addConstraint{a_1 \lor a_2 \lor \ldots \lor a_i\quad}{}{a_1, \ldots, a_i \in \alpha\protect\label{prob:structure-based1}}
\addConstraint{\overline{a_1 \land a_2}}{}{a_1, a_2 \in \alpha, a_1 \neq a_2\protect\label{prob:structure-based2}}
\addConstraint{a_1 \lor a_2 \lor \ldots \lor a_i\quad}{}{f \in fields, a_1, \ldots, a_i \in \beta_f\protect\label{prob:structure-based3}}
\end{mini!}

\subsubsection{Instantiating Reconstruction Plans}
Our goal is to plain-code serialize objects that are observed at runtime. 
We instrument the specific points in the program where we want to serialize an object.
This instrumentation adds instructions that capture values at runtime, which are used to set the meta-variables within the reconstruction plan of the object. 
At runtime, when serialization is requested, the  monitoring instructions are executed, resulting in objects serialized in plain code via instantiated reconstruction plans.
For example, we can add instructions at the beginning of a specific method in order to serialize its receiver and parameters \cite{pankti}. 

Finally, the instantiated reconstruction plans are converted to a plain-code representation by iterating over the actions (\autoref{tab:action-types}) and converting them one-by-one.
A ``call constructor'' action first plain-code serializes each argument of the constructor and then emits a constructor call statement.
Similarly, the arguments to the ``call factory method'' and ``call method'' actions are plain-code serialized  before emitting a method call statement.
``Assign field'' plain-code serializes the value of the field and emits a field assignment statement, while ``use enum constant'' emits a read to the matching constant.
We illustrate this procedure in \autoref{lst:structure-based-example} by reconstructing a \texttt{Monkey} using the first reconstruction plan from \autoref{lst:reconstruction-plans}, which is the most idiomatic.
For our \texttt{Monkey}, this means emitting a code statement that calls the three-argument constructor.
The first argument, \texttt{age}, is a primitive and is directly inserted.
The other two arguments are recursively plain-code serialized, stored as variables and referenced in the generated constructor call.
The ``use enum constant'' action is used when plain-code serializing the \texttt{eyeColor} of the \texttt{Monkey}.
Lastly, the ``use object reference'' defers serialization to the next strategy, trace-based plain-code serialization, and is explained in \autoref{sec:trace-based-strategy}.
The \texttt{habitat} field of the \texttt{Monkey} is handled by this action type.

\begin{lstlisting}[
  language=Java,
  caption={Structure-based plain-code serialization of a \texttt{Monkey} from \autoref{lst:extraction}},
  label={lst:structure-based-example},
  float,
  belowskip=-1em,
  numbers=left,
  morekeywords={enum}
]
// Serialized using the "use enum constant" action type
Color color = Color.BLUE;
// Serialized using the "use object reference" action type
Habitat habitat = /* trace based, see %*\autoref{sec:object-construction-mixing}*) */;
// Serialized using the "call constructor" action type
Monkey monkey = new Monkey(6, color, habitat);
\end{lstlisting}

\subsection{Trace-based Serialization (\textsc{tb})}\label{sec:trace-based-strategy}

Trace-based serialization consists in monitoring an application at runtime, in order to  record an ordered sequence of \emph{event}s happening on a given object.
After the execution finishes, the event sequence is converted to actions per \autoref{tab:action-types} and source code is generated from these actions.

\subsubsection{Capturing Events} An \emph{event} of interest is the occurrence of an operation which mutates an object.
\revisedjss{The considered events are constructor calls (\texttt{ConstructEvent}), field assignments (\texttt{FieldSetEvent}), or impure method calls updating a field.}
Method call events are split into a start (\texttt{MethodStartEvent}) and an end event (\texttt{MethodEndEvent}), allowing other events to happen in-between.
We capture all calls to methods with at least one field assignment.

We denote all potential events to reconstruct an object $o$ as\\ $event\_sequence(o)$.
Based on this notation, we specialize the general object reconstruction formulation, \autoref{eq:object-reconstruction}, for trace-based reconstruction as follows: 
\begin{equation} \label{eq:TB}
evaluate(event\_sequence(o)) \equiv o
\end{equation}

Given a type we want to serialize, we first identify all types needed for its reconstruction because they are used either as fields, constructor parameters, or method parameters.
Then, we instrument the constructors, accessible methods, and field assignments for all these types.
For example, in the \texttt{Habitat} class (\autoref{lst:extraction}), we instrument the constructor on line $28$, as well as the \texttt{grow} method on lines $30-33$.
We show the resulting instrumented code for the \texttt{grow} method in \autoref{lst:instrumentation}.
When the instrumented \texttt{grow} is invoked at runtime, it emits multiple events, ordered by a monotonically increasing counter representing logical time.

\begin{lstlisting}[
  language=Java,
  caption={Instrumentation for monitoring events within the trace-based object reconstruction strategy, the injected code is highlighted in green.},
  label={lst:instrumentation},
  float,
  belowskip=-1em,
  numbers=left,
  morekeywords={enum}
]
public void grow(double newArea) {
%*\fboxsep 1pt \colorbox{green!20}{+  emitMethodStartEvent("Habitat\#grow(double)", this, newArea);}*)
  double newValue = this.area + newArea;
%*\fboxsep 1pt \colorbox{green!20}{+    emitFieldSetEvent(this, this.area, newValue);}*)
  this.area = newValue;
%*\fboxsep 1pt \colorbox{green!20}{+  emitMethodEndEvent("Habitat\#grow(double)");}*)
}
\end{lstlisting}

The first event, \texttt{MethodStartEvent}, marks the start of a method call, and contains all information necessary to convert it to a method call action: the receiver object, the fully qualified name of the invoked method, as well as all arguments to the method call.
As method invocations can be nested, and methods can be recursive, we also track when a method invocation ends through the emission of a \texttt{MethodEndEvent} when the method returns.
The start and end events share a common unique ID, which is used to pair them up.
We also assign unique identifiers to each object.
This ensures that all method invocations, constructor calls, and field assignments for an object can be associated.
The object ID is used whenever an object is referenced within an event.
In addition to the object ID, every reference also stores the current logical time, as an object may have different states at different points in time.
We also see on line $5$ of \autoref{lst:instrumentation} that the \texttt{grow} method updates the value of the \texttt{area} field.
This corresponds to a \texttt{FieldSetEvent} event, which is emitted whenever a field of an object is reassigned inside a method.
\revisedjssold{We assume that the principle of encapsulation is followed, and fields are either all public for simple data classes, or not mutated from outside the class itself.}
Within the \texttt{FieldSetEvent}, we store the ID of the receiver object, as well as the new and old values of the field being updated.
Using \texttt{FieldSetEvent}s, we can determine whether a given method invocation mutated the receiver object, and would need to be replayed during object reconstruction.
Besides capturing method invocations and field assignments, we also instrument all constructors.
Whenever a constructor finishes executing, a \texttt{ConstructEvent} is emitted.
This event contains all necessary information to create a constructor call action, including the fully qualified name of the invoked constructor, as well as all of its arguments.
Additionally, the event also contains the ID of the newly created object, allowing later events to refer to it.
Thus, exercising the instrumented program with a workload emits events.
The resulting captured event sequence is persisted on disk.

\subsubsection{Mapping Events to Actions}\label{sec:tb-event-sequence-analysis}
After execution has concluded, the persisted event sequence is analyzed by iterating over it and mapping each event to an action of \autoref{tab:action-types} if appropriate.
Whenever a \texttt{ConstructEvent} is found, the ID of the created object is saved and a ``call constructor'' action is created.
The same action type is used by the structure based serialization, as shown in \autoref{tab:action-types}.
Every field assignment or method call within the constructor is ignored, as they would already be executed when calling the constructor.
Next, every pair of \texttt{MethodStartEvent} and \texttt{MethodEndEvent} is associated with the receiver and discarded if it did not mutate the receiver.
A method invocation is considered mutating, if a \texttt{FieldSetEvent} occurs on the receiver, or a field within it, before the corresponding \texttt{MethodEndEvent}.
As this requires each object to know exactly what its fields are at each point in time, the initial field values set by the constructor are stored in the \texttt{ConstructEvent} and updated using the encountered \texttt{FieldSetEvent}s.
A ``call method'' action (\autoref{tab:action-types}) is created for each mutating method invocation.
After the whole event sequence is processed, every object should now be associated with at least one constructor call action and zero or more method call actions.

Whenever the plain-code serialization of an object is requested, we follow the same procedure as for structure-based serialization.
We illustrate this by plain-code serializing an instance of the \texttt{Habitat} class from \autoref{lst:extraction}.
The resulting code is shown in \autoref{lst:trace-based-example}.
First, the constructor call action is converted to code by plain-code serializing all constructor arguments.
In our example, this means serializing the \texttt{coordinate} parameter of the constructor.
Next, a constructor call is emitted and the newly created object is stored in a local variable.
This is done on line $2$ of \autoref{lst:trace-based-example}.
Then, all method call actions up until the requested logical time are also converted to code by plain-code serializing the parameters and then calling the method on the created object.
For our \texttt{Habitat} this is a single method call, \texttt{grow(42)}, on line $4$.
\begin{lstlisting}[
  language=Java,
  caption={Trace-based plain-code serialization of a \texttt{Habitat} from \autoref{lst:extraction}},
  label={lst:trace-based-example},
  float,
  belowskip=-1em,
  numbers=left,
  morekeywords={enum}
]
// Constructor call action
Habitat habitat = new Habitat("59.34743,18.07378");
// Method call action
habitat.grow(42);
\end{lstlisting}

\subsubsection{Object Reconstruction Mixing}\label{sec:object-construction-mixing}

At runtime, one may plain-code serialize only with pure structure-based or with pure trace-based serialization.
However, this is suboptimal.
In fact, a single object may be plain-code serialized using a combination of both structure- and trace-based strategies. We call this ``object reconstruction mixing''.

For example, while the primary object  might be serialized with the trace-based strategy, the arguments of the corresponding constructor or method call actions may be serialized using the structure-based strategy.
Likewise, an object  might be serialized using the structure-based strategy, but its fields may require trace-based serialization.

We show such a case in \autoref{lst:mixing-example-monkey}, where we are constructing a \texttt{Monkey}.
\texttt{Monkey} is serialized using structure-based serialization, and its \texttt{habitat} field is serialized using trace-based serialization.
As the captured events are mapped to code offline, but the plain-code serialized \texttt{habitat} field has to be inserted into a specific spot in the generated code, the structure-based serialization strategy uses the ``use object reference'' (from \autoref{tab:action-types}) marker action.
This action emits an $id@logical-time$ reference, which is replaced by the trace-based serialized value after the event sequence conversion is finished.
Besides this action, \autoref{tab:action-types} also contains two more actions, ``use static field'' and ``use standard charset''.
These are actions that can not be done structurally, but also do not need the full trace-based serialization machinery.
For the ``use static field'' action type, we look at all static fields of the class we want to serialize and detect if any of these store a reference-equal object.
If such a field can be found, we emit a field read instead of tracing the constructor and subsequent mutating method calls.
A similar, but more specific, optimization is realized by the ``use standard charset'' action type.
This action type replaces references to Java's standard charsets with the canonical named constants, and serves as an example of integrating domain logic into the serialization process.
To further facilitate extensibility, such custom adapters and action types can be provided as plugins.

\begin{lstlisting}[
  language=Java,
  caption={Plain code serialization of a \texttt{Monkey}, mixing structure- and trace-based serialization},
  label={lst:mixing-example-monkey},
  float,
  numbers=left,
  morekeywords={enum}
]
// Color uses the structure-based strategy
Color color = Color.BLUE;
// Habitat uses the trace-based strategy
Habitat habitat = new Habitat("59.34743,18.07378");
habitat.grow(42);
// Monkey uses the structure-based strategy
Monkey monkey = new Monkey(6, color, habitat);
\end{lstlisting}

\subsection{\proDJ: Plain-code Serialization in Java}\label{sec:prodj}

\begin{figure*}
    \centering
    \includegraphics[scale=0.6]{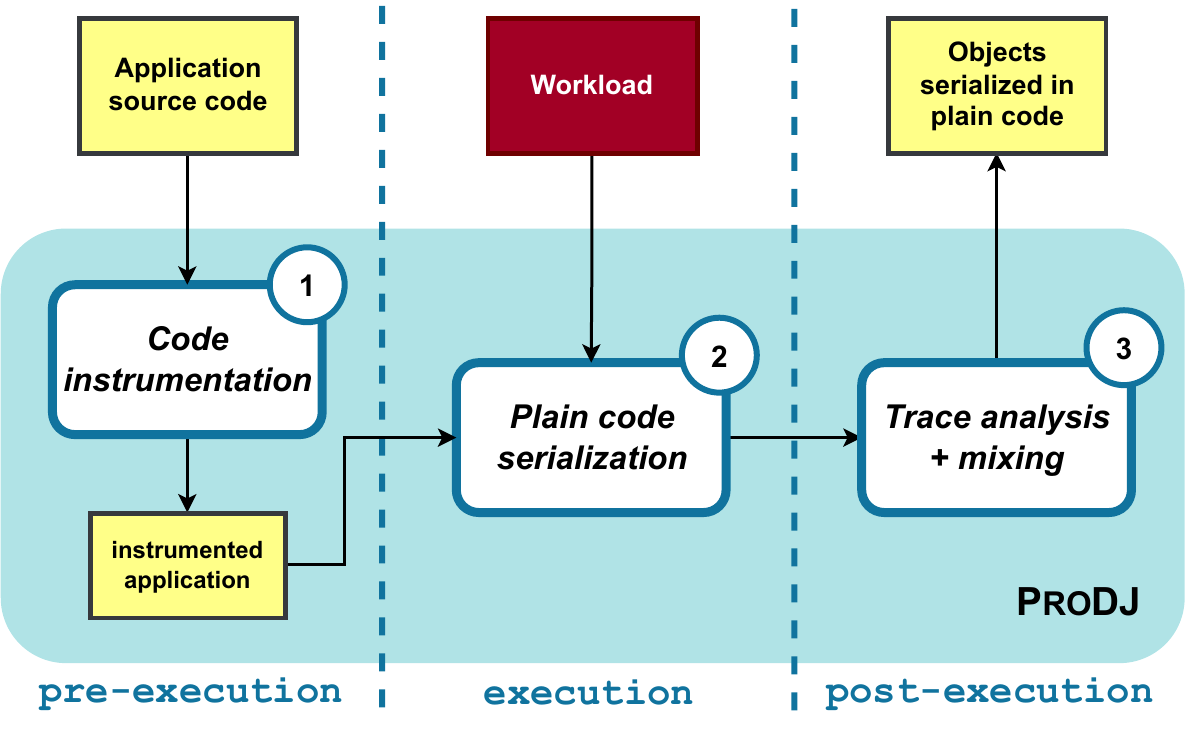}
     \caption{\proDJ uses plain-code serialization for capturing objects at runtime.}
    \label{fig:prodj-overview}
\end{figure*}

We implement a prototype of plain-code serialization in the context of Java, in a tool called \proDJ.
The \proDJ workflow is presented in \autoref{fig:prodj-overview}.
We now describe each of its phases, and also provide details on its implementation.

\subsubsection{Pre-execution}
Per \autoref{fig:prodj-overview}, the first phase within \proDJ occurs offline, before the application is launched.
During this phase, \proDJ takes as input the source of a Java application, and identifies the types associated with each serialization point of interest.
Next, \proDJ instruments each of these points for runtime monitoring, so that the corresponding objects can be plain-code serialized.
For example, instructions can be added at the beginning of a method to plain-code serialize its receiver and parameter objects, including the objects within them.
As detailed previously, \proDJ instruments associated types such that reconstruction plans are synthesized for types eligible for structure-based serialization, and events are captured for associated types that must be trace-based serialized.
This first phase of \proDJ results in an instrumented version of the application, which includes additional instructions required for plain-code serialization.

\subsubsection{Execution}
The second phase of \proDJ occurs online during the execution of the Java application, as illustrated in \autoref{fig:prodj-overview}.
During this phase, the application instrumented by \proDJ is exercised with a workload.
The workload may trigger the instrumentation instructions, and consequently, the plain-code serialization of the corresponding objects.
If an object is serialized using the structure-based strategy, its instantiated reconstruction plan is embedded directly in the log.
For an object serialized using the trace-based strategy, \proDJ logs the sequence of captured events.
To summarize, as the application executes under a workload, \proDJ plain-code serializes the objects that are specified at specific instrumentation points.

\subsubsection{Post-execution}
The third phase of \proDJ occurs offline.
During this phase, \proDJ parses the logs captured in the previous phase, i.e., at runtime.
Recall that objects of interest can be serialized using both structure- and trace-based strategies.
In the former case, \proDJ has already produced plain code instantiated reconstruction plans.
In the latter case, \proDJ maps the captured event sequence to actions.
In both cases, plain-code representations of associated objects are produced from the actions.
Furthermore, \proDJ performs extra steps to enhance the readability and idiomacity of the plain-code objects.
This includes de-duplicating redundant objects and outlining statements into helper methods to set up objects.
The output of this phase of \proDJ is plain Java representations of objects seen in production.

\subsubsection{Implementation}
\proDJ relies on Spoon \cite{pawlak:hal-01169705} for source code analysis, in order to identify the associated types of methods involving serialization within the target Java application.
\proDJ uses Z3 \cite{de2008z3} to solve the constraint problem of \autoref{prob:structure-based} and determine the most idiomatic reconstruction plan for types that are structure-based serializable.
The event sequences for trace-based serializable objects are logged in JSON.
\revisedjssold{The instrumentation of methods and their associated types is achieved using ByteBuddy\footnote{\url{https://bytebuddy.net/}}}.

\section{Experimental Methodology}
\label{sec:evaluation}

This section describes how we empirically evaluate \proDJ, our prototype of plain code serialization for Java.
First, in \autoref{sec:test-generation}, we describe an application of \proDJ for the generation of tests from runtime.
Next, \autoref{sec:case-studies} introduces the real-world projects we use as the subjects for this evaluation.
This is followed in \autoref{sec:workloads} by a description of the workloads we use with each project for our experiments.
Finally, in \autoref{sec:research-questions}, we present the research questions we answer with our evaluation, as well as the protocol we use to do so.

\subsection{Use-Case: Test Generation}\label{sec:test-generation}
In order to evaluate \proDJ, we envision a use case for it. 
We monitor the invocation of a target method within an executing application, use \proDJ to capture objects associated with the method, and then automatically generate tests for it with plain-code objects.
The use of plain-code serialized objects in tests eliminates the dependence on test fixtures and external resources, which contain objects serialized in JSON \cite{alshahwan2024observation} or XML formats \cite{pankti}. 
Instead, generated tests have test inputs that are expressed in the native programming language, while also reflecting actual production states \cite{wang2017behavioral}.

First, \proDJ finds methods within the application that can be suitable candidates for test generation, based on a configurable set of criteria.
We refer to a candidate method as a method under test, or an MUT.
Next, for each MUT, the three phases of \proDJ described in \autoref{sec:prodj} are applied.
The final output of \proDJ is automatically generated JUnit tests\footnote{\url{https://junit.org/junit5/}}, which follow the \emph{Arrange-Act-Assert} pattern.
In contrast to previous work \cite{pankti} where production objects serialized as XML are used in the \emph{Arrange} part, such as in \autoref{lst:introduction-test-with-xml}, or as JSON within resource files \cite{alshahwan2024observation}, \proDJ makes it possible to use plain-code serialized objects.
Within the \emph{Act} phase, the MUT is invoked on the receiver object, with plain-code serialized arguments.
Finally, the \emph{Assert} phase verifies that the output from the invocation of the MUT within the test is the same as the one observed in production.
If the assertion involves an object, the expected value of the assertion is also a plain-code serialized object.

\subsection{Subjects}\label{sec:case-studies}

\begin{table}[]
\begin{tabular}{|lr|r|r|r|r|r|}
\hline
\multicolumn{1}{|l|}{\textsc{Project}} &%
\textsc{SHA} &%
\textsc{LOC} &%
\textsc{\#SPs} &%
\textsc{\#Types} &%
{\begin{tabular}[c]{@{}r@{}}\textsc{\#SPs} \\ \textsc{Prod}\end{tabular}} &%
{\begin{tabular}[c]{@{}r@{}}\textsc{\#Types} \\ \textsc{Prod}\end{tabular}} \\
\hline
\multicolumn{1}{|l|}{\pdfbox} & \textbf{\texttt{\href{https://github.com/apache/pdfbox/tree/8f23f8791c3a526c22cf6a0f6f0d19d830ced0d0}{8f23f8791}}} & 170,823  & 1,390 & 738 & 183 & 188 \\
\hline
\multicolumn{1}{|l|}{\graphhopper} & \textbf{\texttt{\href{https://github.com/graphhopper/graphhopper/tree/f64c57c2ec19c83fa80054bd53a9f08e88b497d3}{f64c57c2e}}} & 96,716 & 945 & 590 & 199 & 114 \\
\hline
\multicolumn{1}{|l|}{\broadleaf} & \textbf{\texttt{\href{https://github.com/BroadleafCommerce/BroadleafCommerce/tree/951216e942b976162dfac048983ee9f7cba888dd}{951216e942}}} & 198,602 & 2,933 & 1,883 & 508 & 444 \\ \hline
\multicolumn{1}{|l|}{\gephi} & \textbf{\texttt{\href{https://github.com/gephi/gephi/tree/82684f92736194a4486ea4840948947489e140a7}{82684f927}}} & 130,705 & 1,057 & 1,721 & 140 & 225 \\
\hline
\multicolumn{3}{|l|}{\textsc{Total}} & 6,325 & 4,932 & 1,030 & 971 \\
\hline
\end{tabular}
\caption{\revisedjss{Projects used in the evaluation of plain code serialization: the exact version (\textsc{SHA}) we use for our experiments, the number of lines of Java code (\textsc{LOC}), the number of serialization points we consider (\textsc{\#SPs}), the number of types related to the serialization points (\textsc{\#Types}), the number of serialization points triggered in production by our workloads (\textsc{\#SPs Prod}), and the number of types that we target for plain code serialization at runtime (\textsc{\#Types Prod}).}}
\label{tab:case-studies}

\end{table}

\revisedjss{We evaluate \proDJ with $4$ open-source Java projects that are popular and active, and span diverse application domains.}
We present them in \autoref{tab:case-studies}.
\pdfbox is a library and command-line tool for the manipulation of PDF documents\footnote{\url{https://pdfbox.apache.org/}}.
Its features include conversion between text and PDF documents, extraction of images, and merging of PDF documents, among others.
\graphhopper is an open-source routing application\footnote{\url{https://www.graphhopper.com/}}, used to compute a path between input locations on a map.
\broadleaf is an e-commerce web application\footnote{\url{https://www.broadleafcommerce.com/}}.
\gephi is a graphical tool for working with graph data\footnote{\url{https://gephi.org/}}, including capabilities for visualization and analysis.

\revisedjssold{For each project in \autoref{tab:case-studies}, column $2$ links to the exact version (\textsc{SHA}) we use for our evaluation, and column $3$ presents the number of lines of Java source code within each project, calculated with \href{https://github.com/XAMPPRocky/tokei}{\texttt{tokei}}}.
Columns $4$ and $5$ give the number of methods instrumented by \proDJ within the project (\textsc{\#SPs} for `serialization point'), and consequently, the number of associated types (\textsc{\#Types}) that need to be instrumented at this point.
\revisedjss{For example, \proDJ instruments $2,933$ methods within \broadleaf, involving $1,883$ types.}
\revisedjss{For our evaluation, we select and instrument methods within each project that are public, non-abstract, non-static, non-deprecated, contain more than one statement, and are declared in a public, non-anonymous, non-local, and non-deprecated class.}
Columns $6$ and $7$ of \autoref{tab:case-studies} provide the same metrics, but computed after runtime observations, i.e., the number of serialization points and types that are covered at runtime, which we discuss in the next subsection.

\subsection{Workloads}\label{sec:workloads}
\revisedjss{The key reason we select the projects presented in \autoref{tab:case-studies} for evaluating plain-code test generation with \proDJ is that we can design representative and repeatable production workloads for each of them, which is essential for our systematic experimental protocols.}
We invoke \pdfbox through its command line interface, and use its ``render'' command to convert a PDF document containing multiple pages into JPEG image files.
We run the \graphhopper server, and query \graphhopper for a route between two coordinates in Berlin\footnote{\url{https://bit.ly/3LA2PVu}}.
In \broadleaf, we add some products to our shopping cart, create a user account, log in, and favorite some products.
Next, we browse through our shopping cart and change the quantity of some products, before finally completing our order.
We use the graphical interface of \gephi to load the ``Les Mis\'erables'' dataset\footnote{\url{https://github.com/gephi/gephi/wiki/Datasets\#social-networks}}, and compute all the statistics on it that \gephi offers.
This includes computing the average degree of nodes, the network diameter, as well as operations like connected components or Pagerank.

Our interactions with each project result in the invocation of serialization points and correspondingly, the instantiation of types that are instrumented.
The last two columns of \autoref{tab:case-studies} give those numbers.
For example, of the $2,933$ methods that are instrumented by \proDJ for \broadleaf, $508$ are triggered by our workload (\textsc{column \#SPs Prod}).
\revisedjss{These methods become the methods under test.}
Consequently, $444$ unique types involved in serialization are instantiated (column \textsc{\#Types Prod}).
As we shall see in \autoref{sec:rq1}, these serialization points will trigger the serialization of thousands of objects under our workload.

\subsection{Research Questions}\label{sec:research-questions}
We exercise the $4$ software applications introduced in \autoref{sec:case-studies} with the production workloads defined in \autoref{sec:workloads}.
\proDJ observes these executions and generates plain-code serialized objects within tests.
We answer the following research questions based on this evaluation.

\subsubsection*{\textbf{RQ1 (Objects)}}
\emph{To what extent can proDJ serialize runtime objects in plain code?}\\
\revisedjssold{In order to answer this RQ, we report the number of plain-code serialized objects for each project under the considered workload (\textsc{\#Obj Plain Code}).}
We also present the number of objects serialized solely through the structure-based strategy (\textsc{sb}), as well as those serialized using the trace-based strategy (\textsc{tb}).
The findings from this RQ illustrate plain-code serialization in action, and highlight the ability of \proDJ in achieving plain-code serialization for real-world Java projects.

\subsubsection*{\textbf{RQ2 (Accuracy)}}
\emph{To what extent does \proDJ successfully reconstruct production objects from plain code?}\\
For this RQ, we use the tests generated (\textsc{\#Tests}) by \proDJ.
In contrast to RQ1, these tests contain assertions that help us assess the correctness of deserialized objects.
We execute all generated tests, and report the number of tests that \textsc{Pass} as well as the number of tests that \textsc{Fail}.
This RQ allows us to identify objects for which plain-code serialization comes with strong consistency verified with assertions.

\subsubsection*{\textbf{RQ3 (Performance)}}
\emph{What is the serialization performance of \proDJ?}\\
To answer this RQ, we compute the overhead for the three phases of \proDJ, pre-execution, execution, and post-execution.
The pre-execution overhead consists of source code analysis and preparation for runtime capturing.
We use the operating system clock to measure the duration of this phase and normalize the duration by dividing it by the number of analyzed types.
Subsequently, at runtime, \proDJ serializes objects and instruments the bytecode of traced classes.
We measure the wall-clock time for serialization and normalize it by the amount of serialized objects.
For the ByteBuddy instrumentation overhead we use \texttt{async-profiler}\footnote{\url{https://github.com/async-profiler/async-profiler}}.
The post-execution overhead corresponds to the wall-clock time taken for trace-analysis for object reconstruction, and actual code generation.
We also measure the time taken to optimize for readability, such as outlining object reconstruction into helper methods, as well as the eventual writing of the generated tests to disk.
The total time is then normalized by the number of created tests.

\subsubsection*{\textbf{RQ4 (User Study)}\label{sec:rq4-protocol}}
\revisedjssold{\emph{What is the opinion of developers on tests containing plain-code serialized objects?}}\\
\revisedjssold{
With this RQ, our goal is to assess the opinion of developers on the representation of serialized objects as plain code, relative to other serialized textual representations, specifically XML and JSON. 
In order to do this, we first select a set of two generated tests for each of the four projects under study, which contain objects serialized at runtime.
Next, we invite 17 developers in our professional network over email, to participate in an online survey.
As part of the survey, we present each developer a set of five random pairs of generated tests.
Each pair contains two variants of the same test, contrasting XML or JSON with \proDJ-produced plain-code serialized objects.
This means that one test in the pair contains either XML or JSON, while the other contains \proDJ-produced plain-code serialized objects.
We invite each participant to mark the test in each pair that is more readable, according to their preference.
For each pair of tests that receives two responses, we compute the Cohen's kappa coefficient for agreement.
For pairs that receive more than two responses, we report the Fleiss' kappa \cite{fleiss1973equivalence}.
We analyze the responses from this survey and present our findings in \autoref{sec:rq4-results}.
Our replication package contains all details of this survey.
}

\section{Experimental Results}\label{sec:results}
This section presents the results from our evaluation of \proDJ with the $4$ projects.
 
\begin{table*}
\renewcommand*{\arraystretch}{1.3}
\footnotesize
\centering
\resizebox{\textwidth}{!}{
\begin{tabular}{|l||rrr||rrr|}
\hline
\multicolumn{1}{|c||}{\multirow{2}{*}{\textsc{Project}}} & 
\multicolumn{3}{c||}{\textsc{RQ1: Objects}} &
\multicolumn{3}{c|}{\textsc{RQ2: Accuracy}} \\ \cline{2-7}
\multicolumn{1}{|c||}{} & 
\multicolumn{1}{c|}{\textsc{\begin{tabular}[c]{@{}c@{}}\#Obj Plain\ \\ Code\end{tabular}}} &
\multicolumn{1}{c|}{\textsc{sb}} &
\multicolumn{1}{c||}{\textsc{tb}} &
\multicolumn{1}{c|}{\textsc{\#Tests}} & 
\multicolumn{1}{c|}{\textsc{Pass}} &
\multicolumn{1}{c|}{\textsc{\begin{tabular}[c]{@{}c@{}}Median \\Obj \\ Per Test\end{tabular}}} \\ \hline
\pdfbox & \multicolumn{1}{r|}{100,830} & \multicolumn{1}{r|}{29,385} & 71,445 & \multicolumn{1}{r|}{489} & \multicolumn{1}{r|}{376 (82\%)} &
\multicolumn{1}{r|}{4}  \\ \hline
\graphhopper & \multicolumn{1}{r|}{40,859} & \multicolumn{1}{r|}{38,325} & 2,534 & \multicolumn{1}{r|}{258} & \multicolumn{1}{r|}{191 (74\%)} &
\multicolumn{1}{r|}{3} \\ \hline
\broadleaf & \multicolumn{1}{r|}{30,273} & \multicolumn{1}{r|}{13,929} & 16,344 & \multicolumn{1}{r|}{904} & \multicolumn{1}{r|}{403 (45\%)} &
\multicolumn{1}{r|}{2} \\ \hline
\gephi & \multicolumn{1}{r|}{2,737} & \multicolumn{1}{r|}{1,422} & 1,315 & \multicolumn{1}{r|}{82} & \multicolumn{1}{r|}{68 (82\%)} & 
\multicolumn{1}{r|}{2} \\ \hline
\textsc{Total} & \multicolumn{1}{r|}{174,699} & \multicolumn{1}{r|}{83,061} & \multicolumn{1}{r||}{91,638} & \multicolumn{1}{r|}{1,733} & \multicolumn{1}{r|}{1,038} &
\multicolumn{1}{r|}{2} \\ \hline
\end{tabular}
}
\caption{\revisedjss{Results of plain-code serialization on $4$ open-source, real-world Java applications}}
\label{tab:results}

\end{table*}

\subsection{RQ1 (Objects)}
\label{sec:rq1}

Columns 2 to 4 of \autoref{tab:results} present the results for RQ1.
First, we note that \proDJ successfully plain-code serializes production objects for all $4$ projects.
For instance, \proDJ plain-code serializes $100,830$ objects for \pdfbox.
Overall, \proDJ plain-code serializes $174,699$ objects that occur at the considered serialization points.
Column 3 highlights that nearly 48\% of the objects ($83,061$) are plain-code serialized using the structure-based strategy.
This means that \proDJ synthesized a minimal reconstruction plan for each of these objects, and instantiated the meta-variables within it, at runtime.
Next, from column 4, we see that $91,638$ objects (52\%) were serialized via the trace-based strategy.
As described in \autoref{sec:trace-based-strategy}, this means that \proDJ either found a static field with the same value, or recorded constructor and mutating method calls that can reconstruct the object to its target state.

To illustrate plain-code serialization, \autoref{lst:prodj-test} shows an excerpt of the serialized version of a \texttt{GHRequest} object in \graphhopper.
This serialized object is concise, and we note that the variable names refer to the semantics of the program.
The names of the variables, such as \texttt{points1}, correspond to the names of the parameters they are passed as, and are used thanks to source code analysis happening together with plain-code serialization.

\texttt{com.graphhopper.routing.SPTEntry}, which makes up the shortest-paths tree in graphhopper, is another good example.
Its entries form a tree structure, each entry has a pointer to its parent, and information about edges, nodes and weights.
\proDJ faithfully recreates the recursive structure, unrolling and wiring up the children.
To do this, \proDJ naturally starts with the root, and then creates the other entries in dependency order.
We see that the proportion of objects that are eligible for the structure- and trace-based strategies is impacted by the application domain. 
\pdfbox uses constants derived from the PDF specification, such as \texttt{Author} and \texttt{Colors}.
These correspond  to static fields, resulting in a majority of objects being eligible for trace-based reconstruction with \proDJ.
On the other hand, \graphhopper defines collections that \proDJ can effectively reconstruct with the structure-based strategy.
From \autoref{tab:results}, we see that each strategy is applied to roughly half of the objects in \broadleaf and \gephi, resulting from the interactions with the UI as part of our workload.

\begin{mdframed}[nobreak=true,style=mpdframe,frametitle=Answer to RQ1]
\proDJ  successfully plain-code serializes thousands of objects ($174,699$) observed at runtime in $4$ real-world Java applications.
Both the structure-based and trace-based serialization strategies are useful to translate runtime objects in plain Java. 
\end{mdframed}

\begin{lstlisting}[
  language=Java,
  caption={Excerpt of the plain-code serialized \texttt{GHRequest} object from \graphhopper, requesting a route between two points. Plain-code serialization uses a combination of the structure- and trace-based strategies.},
  label={lst:prodj-test},
  float,
  numbers=left,
  morekeywords={enum}
]
// Plain-code serialization based on the trace-based strategy
GHRequest object = new GHRequest();
// GHPoint uses the structure-based strategy
GHPoint points = new GHPoint(52.565499, 13.469925);
GHPoint points0 = new GHPoint(52.508281, 13.326416);
// The point list uses the structure-based strategy
List<GHPoint> points1 = new ArrayList<>(List.of(points, points0));
object.setPoints(points1);
%*$\cdots$*) %*\setcounter{lstnumber}{16}*)
object.setCurbsides(curbsides);
\end{lstlisting}

\subsection{RQ2 (Accuracy)}

\begin{lstlisting}[
  language=Java,
  caption={A test generated by \proDJ for \graphhopper},
  label={lst:generated-test-graphhopper},
  float,
  belowskip=-1em,
  numbers=left,
  breaklines=true,
  morekeywords={enum}
]
@Test
@DisplayName("alignOrientation(double,double)")
void test774AlignOrientation$double_double() {
  // Arrange: Recreated with SB
  AngleCalc receiver = new AngleCalc();
  double baseOrientation = -1.971212484045175;
  double orientation = 2.744264992492817;
  // Act
  double actual = receiver.alignOrientation(baseOrientation, orientation);
  // Assert
  assertThat(actual).usingComparator(doubleNanAwareComparator()) .isEqualTo(-3.538920314686769);
}
\end{lstlisting}

\begin{lstlisting}[
  language=Java,
  caption={A test generated by \proDJ for \pdfbox. A part of the test has been outlined to the helper method \texttt{createPDColor}, that recreates production objects},
  label={lst:generated-test-pdfbox},
  float,
  belowskip=-1em,
  numbers=left,
  breaklines=true,
  morekeywords={enum}
]

@Test
@DisplayName("getComponents()")
void test557GetComponents() throws IOException {
  // Arrange
  PDColor receiver = this.createPDColor();
  // Act
  float[] actual = receiver.getComponents();
  // Assert
  assertThat(actual).isEqualTo(new float[]{0.917647F, 0.92549F, 0.937255F});
}

PDColor createPDColor() throws IOException {
  // Recreated with TB
  COSArray receiver = new COSArray();
  COSBase objectsList = COSFloat.get("0.917647");
  COSBase objectsList0 = COSFloat.get("0.92549");
  COSBase objectsList1 = COSFloat.get("0.937255");
  // Recreated with SB
  Collection<COSBase> objectsList2 = new ArrayList<>(List.of(objectsList, objectsList0, objectsList1));
  // Recreated with TB
  receiver.addAll(objectsList2);
  PDColorSpace receiver0 = PDDeviceRGB.INSTANCE;
  return new PDColor(receiver, receiver0);
}
\end{lstlisting}


In this research question we reason about the behavior of deserialized objects by running the tests generated by \proDJ.
Columns 5 through 8 of \autoref{tab:results} present the results for RQ2.
\proDJ generates $1,733$ test cases in total for our $4$ study subjects. 
Those test cases contain assertions that enable us to assess the validity of deserialized objects.

Per RQ1, the total number of objects plain-code serialized by \proDJ is $174,699$ but we use only a fraction of them in the $1,733$ generated tests.
We de-duplicate identical objects, before including them within the generated tests.
\revisedjssold{We consider an object to be identical to another if the generated statements are the same, modulo identifier names, or refer to the same object (\texttt{==}) reconstructed using trace-based serialization.}
Furthermore, as trace-based serialization indiscriminately records all instances of the types we monitor, we serialize many objects only for trace-based serialization events.
\revisedjssold{For example, we might record multiple \texttt{GHRequest} instances, but only one is actually used in a method under test.
We then discard all other instances that are not associated with it.}
This means that we do not use all of the $174,699$ plain-code serialized objects within the $1,733$ generated tests.
We present the median number of objects that are deserialized in each each generated test, in the last column of \autoref{tab:results}.
For example, $55$ tests generated for \pdfbox use $1$ plain-code serialized object, while one generated test uses $122$ deserialized objects, resulting from the reconstruction of a \texttt{List} object and its constituent elements.
Overall, the median number of objects in a test generated for \pdfbox is $4$.

In \autoref{lst:generated-test-graphhopper}, we present one test generated by \proDJ for \graphhopper, clearly following  the \emph{Arrange-Act-Assert} pattern.
This test verifies the behavior of the \texttt{alignOrientation} method of the \texttt{AngleCalc} class in \graphhopper.
\proDJ generates the Java statements from lines $5$ to $7$ at runtime using the structure-based strategy.
Note that \proDJ reuses the parameter names of \texttt{alignOrientation} for the local variables in the test on lines $6$ and $7$.
The assertion on line $11$ verifies that the \texttt{double} value output from the invocation of \texttt{alignOrientation} on line $9$ matches the value recorded at runtime.
We also present one \proDJ-generated test for \pdfbox in \autoref{lst:generated-test-pdfbox}.
The method under test, called \texttt{getComponents} defined in the class \texttt{PDColor}, returns a non-null, \texttt{float} array output, based on the value of a field called \texttt{colorSpace}.
Note that \proDJ outlines the recreation of the receiver object of type \texttt{PDColor} into a separate \texttt{createPDColor} method.
Within this method, \proDJ uses object reconstruction mixing for plain-code serializing \texttt{PDColor}.
Specifically, during the pre-execution phase, \proDJ discovers that the associated type \texttt{COSArray} is eligible for trace-based serialization.
At runtime, \proDJ records a constructor call event for \texttt{COSArray} (recreated on line $15$), followed by a call to the method \texttt{addAll} (line $22$).
The only parameter of \texttt{addAll} is a list, which \proDJ serializes using the structure-based strategy (line $20$).
However, the contents of this list are serialized using the trace-based strategy (lines $16$ - $18$).
\proDJ uses the parameter name \texttt{objectList} of \texttt{addAll} to derive the names of the local variables in the helper method.
Note that \proDJ observes the calls to the factory method \texttt{COSFloat.get} at runtime and therefore recreates these calls, instead of invoking the \texttt{COSBase} constructor directly.
The \texttt{String} arguments for these \texttt{COSFloat.get} method calls are directly inlined, in contrast with the \texttt{double} arguments on lines $6$ and $7$ in the first test of \autoref{lst:generated-test-graphhopper}.
This is because \proDJ tries to inline primitive values, but only if this would not break up the \emph{Arrange-Act-Assert} pattern.
Next, on line $23$ of \autoref{lst:generated-test-pdfbox}, a \texttt{PDColorSpace} object is reconstructed using the runtime trace as being an instance of the RGB color space.
Finally, the two-argument constructor call on line $24$ reuses the reconstructed objects to recreate the runtime \texttt{PDColor} object.
This object is used within the generated test as the receiver for the invocation of \texttt{getComponents} (line $8$).
Note also the inline creation of the \texttt{float} array on line $10$ within the assertion, which verifies the equality of the expected and actual \texttt{float} array output.
We argue that these plain-code serialization and readability optimization strategies employed by \proDJ result in  tests that not only reflect production observations, but are also focused, well-structured, an easy-to-read.
\revisedjss{Each generated test focuses on the invocation of a single target method on the reconstructed receiver object. 
Consequently, the tests generated by \proDJ do not contain test smells: they are not `eager tests' and do not contain multiple assertions \cite{garousi2018smells}. We note that these tests may miss inconsistencies within other fields of the receiver that are not visible from the assertion.
}

\revisedjssold{All test cases generated by \proDJ compile correctly and when we execute them, $1,038$ test cases pass, which is a success rate of $59.9$\%.
Specifically, for three of the four case studies, a large majority of the generated tests pass.
For \pdfbox, $82$\% of all tests generated pass, the highest among the four projects.}
That is a strong property.
Given the structure of generated tests, it means that the state of an object after calling a method, i.e., after the `Act' phase of the test, is consistent with the state observed in production (the `Assert' phase).
While not being a correctness proof, this is strong consistency evidence.

For \broadleaf, we note that field dependency injection through the Spring framework results in partially initialized objects.
These do not properly reflect production state, causing the test cases to fail.
The failing tests are excellent indicators of the fundamental challenges of serialization, as discussed in \autoref{sec:zoo-of-caveats}.

\begin{mdframed}[nobreak=true,style=mpdframe,frametitle=Answer to RQ2]
\revisedjssold{Our experiment with using plain-code serialization for testing shows that \proDJ faithfully recreates runtime states for all objects in nearly 60\% of the generated tests.}
The assertions in this testing context provide soundness evidence. 
\end{mdframed}

\subsection{RQ3 (Performance)}

The results for RQ3 are presented in \autoref{tab:performance-results}, with the three column groups representing the three measurement categories, corresponding to each of the \proDJ phases discussed in \autoref{sec:prodj}.
\revisedjssold{We run each experiment three times and present the median measurement from the three runs.}
Before the application is run, \proDJ analyses the source code. The time to perform this step is  presented in the column \emph{Pre-execution}.
This includes parsing the source code, building an AST, and extracting all necessary data for structure- and trace-based serialization.
We compute the values by taking the total runtime of this step and dividing it by the amount of types in the project.
For example, during the pre-execution phase, \proDJ analyzes $1,883$ types within \broadleaf, spending $11$ ms per type.
On the other hand, \proDJ spends $27$ ms on each of the $590$ types in \graphhopper.
This difference is mostly because the analysis overhead per type is less than the parsing overhead.
As \broadleaf is a bigger project, the cost is amortized over a larger number of diverse types.

\begin{table*}
\renewcommand*{\arraystretch}{1.3}
\footnotesize
\centering
\resizebox{\textwidth}{!}{
\begin{tabular}{|l|rr|rr|rr|}
\hline
\multicolumn{1}{|c|}{\multirow{2}{*}{Project}} & \multicolumn{2}{c|}{Pre-execution} & \multicolumn{2}{c|}{Execution} & \multicolumn{2}{c|}{Post-execution} \\ \cline{2-7} 
\multicolumn{1}{|c|}{} & \multicolumn{1}{r|}{\begin{tabular}[c]{@{}r@{}}Analysis\\ (ms / type)\end{tabular}} & \begin{tabular}[c]{@{}r@{}}Total time\\ (sec)\end{tabular} & \multicolumn{1}{r|}{\begin{tabular}[c]{@{}r@{}}Serialize\\ (ms / object)\end{tabular}} & \begin{tabular}[c]{@{}r@{}}Total time\\ (sec)\end{tabular} & \multicolumn{1}{r|}{\begin{tabular}[c]{@{}r@{}}Generate\\ (ms / test)\end{tabular}} & \begin{tabular}[c]{@{}r@{}}Total time\\ (sec)\end{tabular} \\ \hline
\pdfbox & \multicolumn{1}{r|}{18} & 14 & \multicolumn{1}{r|}{0.14} & 35 & \multicolumn{1}{r|}{105} & 59 \\ \hline
\graphhopper & \multicolumn{1}{r|}{27} & 16 & \multicolumn{1}{r|}{0.40} & 41 & \multicolumn{1}{r|}{65} & 40 \\ \hline
\broadleaf & \multicolumn{1}{r|}{11} & 23 & \multicolumn{1}{r|}{0.78} & 157 & \multicolumn{1}{r|}{38} & 40 \\ \hline
\gephi & \multicolumn{1}{r|}{11} & 21 & \multicolumn{1}{r|}{0.51} & 159 & \multicolumn{1}{r|}{145} & 29 \\ \hline
\end{tabular}
}
\caption{\revisedjssold{Performance measurements of the different overhead categories}}
\label{tab:performance-results}
\end{table*}

After the analysis is completed, we start the application with \proDJ attached.
The measurements, made when the application is running, are included within the group of columns labeled \emph{Execution}.
Column four reports the average serialization duration per serialized object.
We see that the serialization duration is relatively similar for all projects, with \graphhopper being slightly more expensive.
This is due to the fact that the serializer looks up many nested types for \graphhopper, which requires a full AST traversal.

The last two columns of \autoref{tab:performance-results} present the time taken by \proDJ for post-production.
This duration includes the time spent offline, analysing the event sequence for objects using trace-based serialization in order to reconstruct them, as well as the time taken taken for readability optimizations, and writing of the post-processed generated tests to disk.
The post-execution phase takes under a minute for all projects.
Of the four projects, \proDJ spends most time generating tests for \pdfbox because the tests contain many mutator calls that are made on objects in \pdfbox.
Furthermore, the arguments to these calls often consist of structure-based serialized arrays which have multiple entries.
As \proDJ needs to build an AST representation of a variable's type in order to de-duplicate objects, this necessitates a lot of string parsing.
The time per test is the highest for \gephi, as many tests \proDJ tries to generate contain objects it is unable to fully reconstruct.
Consequently, the test has to be discarded, increasing the time per successfully generated test.

In addition to the performance data presented in \autoref{tab:performance-results}, we use a profiler for a finer-grained analysis of the time spent by \proDJ on its various runtime tasks.
According to our profiles, the class instrumentation makes up around 8\% of the \pdfbox runtime, but around 72\% of the \broadleaf runtime.
Most of the time for \broadleaf is spent deciding whether to instrument a class, as resolving the supertype hierarchy thousands of times per class is not efficient.
Recall that \proDJ uses plain JSON for event sequences and serializing millions of small objects to JSON and writing them to disk is expensive, so \proDJ offloads this to a separate thread.
Whenever an event needs to be written to disk, it is appended to a threadsafe queue instead, which is then drained in batches by a dedicated thread.
Whenever the queue is full, for example, when a multi-threaded application calls instrumented methods in tight loops, the application will be temporarily suspended until the writing thread has caught up.
This minimizes the impact on the running application.
Analyzing the profiler recordings for \pdfbox showed that, while 3.5\% of the time was spent writing the event sequence on different threads, only 0.23\% of the runtime was spent in application code persisting events.
Similarly, around 6\% of the runtime for \graphhopper was spent writing the event sequence in a different thread, and also only 0.12\% of the runtime was spent in application code persisting events.
For \broadleaf, \proDJ instruments 1,197 types, amounting to a total of over 22,000 rewritten field accesses or method calls.
\proDJ uses ByteBuddy and its \texttt{MemberSubstitution} for this, leading to about 72\% of the runtime being spent in bytecode rewriting.

Overall, serializing an object using \proDJ only takes a few milliseconds.
Most of the time spent instrumenting the bytecode outside of this can be attributed to inefficiencies in the usage of ByteBuddy and is not a conceptual limitation.
Similarly, most of the post-processing time is spent building an AST and converting strings to types.

\begin{mdframed}[nobreak=true,style=mpdframe,frametitle=Answer to RQ3]
The performance evaluation of \proDJ shows that plain-code serialization takes a few milliseconds per serialized object, which is invisible to the user in interactive use-cases.
Contrary to traditional serialization techniques, plain-code serialization involves a computation cost before and after execution.
\end{mdframed}

\revisedjssold{\subsection{RQ4 (User Study)}
\label{sec:rq4-results}
We conduct a user study per the protocol defined in \autoref{sec:rq4-protocol}, based on an online survey with developers.
The goal of the survey is to assess the opinion of developers regarding the readability of tests containing plain-code serialized objects.
\revisedjss{This survey received 17 responses from developers who have between 8 and 43 years of experience with programming (median 20), and between 1 and 30 years programming with Java (median 13).}
We prepare 16 pairs of tests in total, presenting five random pairs to each respondent.
We select two tests for each project which have less than 30 lines of Java code, including outlined utility methods.
\revisedjss{From \autoref{tab:readability-results-kappa}, we see that one test each in \pdfbox (\textsc{PDF1}) and \graphhopper (\textsc{GH1}) have 23 lines of code (LOC), the maximum among the 8 plain-code tests generated by \proDJ.}
For the JSON and XML variants of these tests, we collapsed identical lines in the JSON and XML code.
\revisedjss{The maximum number of lines in the tests with JSON and XML tests is 1,794 and 1,784, respectively (both for \textsc{PDF2} in \autoref{tab:readability-results-kappa}).
Note from the table that in each of the 16 test pairs, as well as across all the pairs, the number of lines of code for the plain-code test is significantly lesser than its corresponding JSON and XML representation.}
Each respondent was shown five randomly selected pairs.
One test within the pair contains plain-code serialized objects, while the other was the same test, but with either JSON- or XML-serialized objects. 
We randomized the order in which the plain-code and JSON or XML tests were presented.
The respondents were requested to mark which test in each pair they preferred, with respect to its readability.
Moreover, 11 of the 17 respondents shared additional qualitative comments about serialization.
}

\begin{table*}
\footnotesize
\renewcommand*{\arraystretch}{1.3}
\centering
\resizebox{\textwidth}{!}{
\begin{tabular}{|l|r|rrr|rr|r|}
\hline
\multirow{2}{*}{Project} & 
\multirow{2}{*}{Test pair} & 
\multicolumn{3}{c|}{LOC} & 
\multicolumn{2}{c|}{Votes} & 
\multirow{2}{*}{\begin{tabular}[c]{@{}r@{}}C or F kappa\\ (agreement)\end{tabular}} \\ \cline{3-7}
& 
& 
\multicolumn{1}{r|}{Plain-code} & 
\multicolumn{1}{r|}{JSON} & 
XML & 
\multicolumn{1}{r|}{Plain-code} & 
JSON/XML & 
\\ \hline
\multirow{4}{*}{PDFBox} & 
PDF1--JSON & 
\multicolumn{1}{r|}{\multirow{2}{*}{23}} &
\multicolumn{1}{r|}{76} &
- & 
\multicolumn{1}{r|}{8} & 
0 & 
F 1.0 (perfect) \\ \cline{2-2} \cline{4-8}
& 
PDF1--XML & 
\multicolumn{1}{r|}{} &
\multicolumn{1}{r|}{-} &  
65 & 
\multicolumn{1}{r|}{3} & 
0 & 
F 1.0 (perfect) \\ \cline{2-8}  
& 
PDF2--JSON & 
\multicolumn{1}{r|}{\multirow{2}{*}{8}} & 
\multicolumn{1}{r|}{1,794} &  
- & 
\multicolumn{1}{r|}{4} & 
2 & 
F -0.07 (poor) \\ \cline{2-2} \cline{4-8}
& 
PDF2--XML & 
\multicolumn{1}{r|}{} &
\multicolumn{1}{r|}{-} &  
1,784 & 
\multicolumn{1}{r|}{6} & 
1 & 
F 0.43 (moderate) \\ \hline
\multirow{4}{*}{GraphHopper} & 
GH1--JSON & 
\multicolumn{1}{r|}{\multirow{2}{*}{23}} &
\multicolumn{1}{r|}{349} & 
- & 
\multicolumn{1}{r|}{7} & 
1 & 
F 0.5 (moderate) \\ \cline{2-2} \cline{4-8}
&
GH1--XML & 
\multicolumn{1}{r|}{} &
\multicolumn{1}{r|}{-} & 
259 & 
\multicolumn{1}{r|}{6} & 
1 & 
F 0.43 (moderate) \\ \cline{2-8} 
& 
GH2--JSON & 
\multicolumn{1}{r|}{\multirow{2}{*}{18}} & 
\multicolumn{1}{r|}{192} &  
- & 
\multicolumn{1}{r|}{2} & 
0 & 
C 1.0 (perfect) \\ \cline{2-2} \cline{4-8}
& 
GH2--XML & 
\multicolumn{1}{r|}{} & 
\multicolumn{1}{r|}{-} & 
147 & 
\multicolumn{1}{r|}{6} & 
0 & 
F 1.0 (perfect) \\ \hline
\multirow{4}{*}{BroadLeaf} & 
BL1--JSON & 
\multicolumn{1}{r|}{\multirow{2}{*}{8}} & 
\multicolumn{1}{r|}{24} &  
- & 
\multicolumn{1}{r|}{4} & 
1 & 
F 0.2 (slight) \\ \cline{2-2} \cline{4-8}
& 
BL1--XML & 
\multicolumn{1}{r|}{} & 
\multicolumn{1}{r|}{-} &  
22 & 
\multicolumn{1}{r|}{9} & 
0 & 
F 1.0 (perfect) \\ \cline{2-8} 
& 
BL2--JSON & 
\multicolumn{1}{r|}{\multirow{2}{*}{10}} & 
\multicolumn{1}{r|}{20} &  
- & 
\multicolumn{1}{r|}{5} & 
0 & 
F 1.0 (perfect) \\ \cline{2-2} \cline{4-8}
& BL2--XML & 
\multicolumn{1}{r|}{} & 
\multicolumn{1}{r|}{-} & 
17 & 
\multicolumn{1}{r|}{2} & 
0 & 
C 1.0 (perfect) \\ \hline
\multirow{4}{*}{Gephi} & 
GEP1--JSON & 
\multicolumn{1}{r|}{\multirow{2}{*}{11}} & 
\multicolumn{1}{r|}{1,128} &  
- & 
\multicolumn{1}{r|}{4} & 
0 & 
F 1.0 (perfect) \\ \cline{2-2} \cline{4-8}
& 
GEP1--XML & 
\multicolumn{1}{r|}{}  & 
\multicolumn{1}{r|}{-} & 
704 & 
\multicolumn{1}{r|}{6} & 
0 & 
F 1.0 (perfect) \\ \cline{2-8} 
& 
GEP2--JSON & 
\multicolumn{1}{r|}{\multirow{2}{*}{11}} & 
\multicolumn{1}{r|}{1,128} & 
- & 
\multicolumn{1}{r|}{3} & 
0 & 
F 1.0 (perfect) \\ \cline{2-2} \cline{4-8}
&
GEP2--XML & 
\multicolumn{1}{r|}{} & 
\multicolumn{1}{r|}{-} & 
704 & 
\multicolumn{1}{r|}{4} & 
0 & 
F 1.0 (perfect) \\ \hline
\textsc{Total} & 
16 & 
\multicolumn{1}{r|}{112} & 
\multicolumn{1}{r|}{4,711} & 
3,702 & 
\multicolumn{1}{r|}{79} & 
6 &  
\\ \hline
\end{tabular}
}
\caption{\revisedjss{Summary of our user study.
We select 2 tests from each of the four projects for the study.
The second column lists the 16 test pairs we prepare using these tests.
Each pair contains one test with plain-code serialized objects and an equivalent test with either JSON or XML representations of the same serialized objects.
The number of lines of code in each test in the pair is indicated in the column LOC.
We present 5 random test pairs to each of the 17 participants of the study, who vote for tests with plain-code (column 3) or JSON/XML (column 4) serialized objects, based on their readability.  
The last column presents the Cohen's (C) or Fleiss' (F) kappas of agreement between the participants.
}}
\label{tab:readability-results-kappa}
\end{table*}

\revisedjssfinal{\autoref{tab:readability-results-kappa} summarizes the responses from the survey.
We find that through 79 of the total 85 votes (92.9\%), the developers showed a preference for the tests containing the plain-code serialized objects, regardless of the differences in the number of lines of code between the plain-code and JSON/XML variants.}
In only 6 cases, developers preferred the versions with XML or JSON representations. 
For 11 of the 16 test pairs, the preference for plain-code serialized objects was unanimous.
In those cases, the agreement measure (Cohen's and Fleiss' kappas) is a perfect 1.0.
In the 5 remaining cases, the agreement varies from poor to moderate.

\revisedjssold{Qualitative statements by the developers confirmed the preference, as the follows:}

\formattedquote{\revisedjssold{I [prefered] the shorter versions because the longer ones contained serialization "noise" that did not seem relevant.}}

\formattedquote{\revisedjssold{Intuitively, the [plain-code version] is often more readable. However, legacy code may be hard to test as classes could have many dependencies that are hard to construct. In those cases, the readability gap between object construction and deserialization could be narrowed.}}

\revisedjssold{A single developer voted for the XML or JSON variants of all presented pairs, hence they are responsible for 5 of the 6 votes for JSON/XML, owing to the higher likelihood of capturing unintentional breakages:}

\formattedquote{\revisedjssold{I prefer the unchanging nature of JSON and XML, which do not get refactored automatically in sync with the application code, when the classes under test change.}}

\revisedjssold{The test pair PDF2--JSON has the lowest agreement score.
This is due to the aforementioned developer who prefers JSON/XML, and another developer who misunderstood the meaning of the test case, per their comment.
Note that the latter respondent ranked plain-code tests higher in the other four pairs they were shown.
}

\revisedjssold{One developer correctly noted that the tests would be organized differently in practice:}

\formattedquote{\revisedjssold{I believe I would store the long text-snippets as variables in another file, instead of having them in the test.}}

\revisedjssold{Finally, a respondent balances the requirements of serialization performance versus readability.:}

\formattedquote{\revisedjssold{When sending data back and forth over the wire, more efficient formats may be better, like Protobuf [...]. Overall, it really depends on the task at hand.}
}

\revisedjssold{This is valid point.
Our technique is entirely optimized for readability, and we do recommend binary serialization for performance-sensitive use cases.}


\begin{mdframed}[nobreak=true,style=mpdframe,frametitle=\revisedjssold{Answer to RQ4}]
\revisedjssold{Our user study demonstrates, through a sound protocol, that developers rate tests containing plain-code serialized objects higher with respect to their readability, in comparison with equivalent tests with XML- or JSON-serialized objects.}
\end{mdframed}

\section{Limitations and Threats}
\label{sec:zoo-of-caveats}

During our prototyping, we have encountered fundamental and interesting challenges of serialization, which we now discuss.

\subsection*{Internal classes}
Serialization may fail because of JDK-internal classes, such as those loaded by the bootstrap-classloader.
When a bootstrap class is instrumented, additional code is inserted and the bootstrap-classloader is used to look up any required \proDJ classes.
As the bootstrap-classloader only loads JDK classes, this does not work and the application crashes.
It would be possible to include our agent in the bootstrap-classpath, but this sophistication is out of the scope of a research prototype.

\subsection*{Streams}
Input streams are problematic as they often read from an ephemeral source, such as a network stream, which might no longer exist when the object is reconstructed.
A simple case illustrating this problem is file access or byte array input streams, as can be seen with \pdfbox.
In this case, \proDJ fails to serialize \texttt{java.io.ByteArrayInputStream}s and \texttt{org.apache.fontbox.afm.CharMetric}.

\subsection*{Runtime code generation}
Runtime annotation processing is not handled by \proDJ.
In Java, annotations are used for dynamic analysis, and for generating implementations or proxy classes on-the-fly.
These classes do not exist on disk, and can therefore not be plain-code serialized.
\revisedjssold{Likewise, for projects such as \broadleaf that heavily rely on dependency injection through runtime code generation, \proDJ can not faithfully recreate all objects.}

\subsection*{Bounded depth} 
\revisedjssold{\proDJ uses a bounded serialization approach to limit the size of objects and consequently the size of generated tests.}
In our experiments, we limit the length of serialized arrays to $25$, in order to have readable test cases.
Consequently, when a test case asserts behavior that requires more than $25$ elements, the assertion fails.
For example, $37$ test cases generated for \pdfbox fail for this reason. 

\subsection*{Visibility}
\revisedjssold{The next challenge relates to visibility and encapsulation, as some serialized fields and methods may not be visible in source code.
For example, the trace-based reconstruction strategy can capture calls to package-private methods, which cannot be replayed in the test.
We observed such a limitation for $22$ tests of \pdfbox.
To address this limitation, \proDJ would need to determine and only use accessible method creating objects of a given type.
A similar limitation occurs when fields are set through reflection or native code.
Such reconstructions are not supported by \proDJ.}

\revisedjssold{\subsection*{Security Implications}
Untrusted object deserialization poses a significant security risk that has been increasingly prevalent in recent years \cite{sayar2023depth}.
This holds for plain-code representations of runtime objects observed in the field.
An attacker could potentially craft a special object, on which they execute a specific malicious sequence of method calls on the testing or CI machine, which is then captured and replayed by \proDJ.
Developers should therefore be mindful of the risks associated with deserialization of production objects.
On the positive side, we note that plain code is easier to audit for security than alternative serialization formats. 
}

\subsection*{State}
\revisedjssold{The last major challenge for serialization relates to identify all parts of the state that are related to the execution.
For example, $11$ tests in \pdfbox fail because they try to read a certain file, which does not exist anymore during test execution.
\proDJ only recreates states that are local to serialized objects, not the complete \emph{system state}.
This is also a potential challenge at reconstruction time, reconstructing a serialized object may incur side effects that we do not handle, such as writing on disk.}

\revisedjssold{\subsection*{Threats to Validity}
Threats to the \emph{internal} validity of our findings may arise from the tools that we use to implement \proDJ.
We mitigate this threat by relying on open-source state-of-the-art tools to handle instrumentation and code transformation.}

\revisedjssold{\emph{External} threats to the validity of our findings are related to the artifacts we use for our experiments.
We demonstrate the feasibility of plain-code serialization of runtime objects through four large, real-world Java projects exercised with representative field workloads. 
We envision that our approach will also be applicable to other Java projects.
}

\revisedjssfinal{A threat to the validity of our conclusions from the \emph{user study} of RQ4 arises from the expectation that developers may automatically prefer shorter versions of tests with plain-code serialized objects over their XML or JSON counterparts.
We design our user study to minimize this threat and also report the deviations we observe from this outcome.
Our key finding is that, even for test pairs with comparable number of lines of code, developers prefer plain-code serialized objects over JSON or XML.}

\revisedjssold{We note that the benefits of plain-code serialization may not \emph{generalize} to all use cases of serialization.
For example, inter-process communication involves the transmission of serialized objects that are not necessarily required to be human-readable.
Serializing these objects in plain code is also likely to be impractical due to performance costs.
On the other hand, where readability is in fact the use case, we believe that our strategy can be adapted to handle plain-code serialization in other programming languages.
The key requirement is that their runtime should support introspection, which is the case for most modern languages, including JavaScript and Python.
}


\section{Related Work}\label{sec:related_work}

This section summarizes studies that propose strategies for more efficient serialization, and those that use serialization in the context of testing. 

The performance optimization of the (de)serialization mechanism has received much attention in the literature.
Runtime code generation can be used to produce code that is tailored to the type being serialized.
This has been proposed for programs written in Java \cite{DBLP:conf/dsn/BennaniBCFKMT04,aktemur2005optimizing, horvath2017code}, C++ \cite{tansey2008efficient}, and Scala \cite{miller2013instant}, and for high-performance computing \cite{tseng2016morpheus}. 
In contrast to these studies targeting performance, \proDJ seeks to optimize readability by expressing objects in idiomatic plain code that is tailored to the project source.

Several studies focus on the construction of an object for use within generated tests.
Some approaches solve this problem by suggesting method sequences for setting up test objects.
MSeqGen \cite{thummalapenta2009mseqgen} mines source code to propose method call sequences on objects associated with a method under test.
These sequences are used within tests generated by Randoop \cite{PachecoLEB2007} or PEX \cite{tillmann2008pex} to increase branch coverage.
OCAT \cite{jaygarl2010ocat} serializes objects and objects captured by OCAT are recreated through a randomly generated method call sequence within tests generated automatically by Randoop \cite{PachecoLEB2007}.
Seeker \cite{thummalapenta2011synthesizing} uses a combination of static analysis and dynamic symbolic execution to produce sequences of method invocations that initialize test inputs to desired states.
Bach \textit{et al.} \cite{bach2020determining} use static analysis to solve the object creation problem.
Their approach suggests optimal method call sequences for the creation of test inputs in C++ programs, but does not address the initialization of objects to a target state.

Serialized representations of objects are also directly used within tests.
Artzi and colleagues propose Recrash \cite{artzi2008recrash}, a technique that maintains a shadow stack to store parts of receiving objects and arguments.
These partial states are serialized and used to generate unit test cases that can reproduce program crashes.
Elbaum \textit{et al.} \cite{elbaum2008carving} serialize objects associated with invoked MUTs during the execution of system tests, in order to use them within generated unit tests.
Pankti \cite{pankti} serializes objects associated with MUTs, while monitoring applications in production.
The objects are deserialized within generated unit tests to recreate production states.
These studies rely on the popular XStream library for serialization to XML, as opposed to plain-code.
TestGen by Meta \cite{alshahwan2024observation} serializes runtime objects for the Instagram application, representing them as JSON for use within unit tests.
Our key contribution is that we serialize runtime objects in plain-code, combining the complementary structure- 
and trace-based strategies.
In the technological space, related to our work is a Python tool called fickling \cite{fickling}.
Instead of directly evaluating a pickled data stream, which could contain malicious executable code, fickling decompiles it into human-readable Python code. 
This is essential, since security is an important aspect at deserialization time \cite{DBLP:conf/icse/GauthierB22,DBLP:journals/tosem/SayarBBT23}.

\revisedjssold{Closely related to our work is CoDeSe, proposed by Gligoric \textit{et al.} \cite{gligoric2011codese}.
CoDeSe, short for COde-based DEserialization and SErialization, is an approach to serialize objects as executable Java code focusing on performance.
While sharing the same idea of plain-code serialization, our work is very different in the outcome.
The objective for serialization with CoDeSe is fast deserialization, and to achieve this, it heavily invokes the Java reflection API and the \texttt{Unsafe} class \cite{OraclePostUnsafe} to reconstruct an object and set its fields.
We invite the reader to peruse Figure 3b of \cite{gligoric2011codese}, which presents the serialized representation of a red-black tree object produced by CoDeSe.
The generated code includes calls to \texttt{unsafe.allocateInstance(...)} which directs the JVM to create the object in memory, as well as calls to \texttt{unsafe.putInt(...)} and \texttt{unsafe.putObject(...)} to manipulate its fields.
CoDeSe can also generate C code to reconstruct objects, which is then processed through the Java Native Interface (JNI).
We argue that this use of reflection and the JNI can be considered as low-level reconstruction.
On the contrary, our approach never uses reflection and unsafe calls, and only uses application-specific constructors and methods that are visible in the deserialization scope.
The focus of plain-code serialization with \proDJ is idiomacity and readability, as illustrated by the serialized objects embedded within the test presented in \autoref{lst:generated-test-pdfbox}.
}

\section{Conclusion}\label{sec:conclusion}

In this paper, we have presented the novel concept of plain-code serialization.
The key benefits are two-fold.
First, the serialized objects are readable by human developers and can be used, for example, in generated tests.
Second, deserialization becomes as trivial as executing the statements of the plain-code serialization.
We have fully implemented the vision in Java, and used our prototype implementation called \proDJ to serialize more than a hundred thousand objects observed at runtime in four real-world Java applications.
Future work can explore the use of advanced impurity analysis to further optimize both structure-based and trace-based plain-code serialization.

\section*{Acknowledgements}
\noindent This work has been partially supported by the Wallenberg Autonomous Systems and Software Program (WASP) funded by the Knut and Alice Wallenberg Foundation.

\bibliographystyle{elsarticle-num} 
\bibliography{main}

\end{document}